# Plasmonics in heavily-doped semiconductor nanocrystals

Francesco Scotognella*[1,3], Giuseppe Della Valle[1,2], Ajay Ram Srimath Kandada[1,3], Margherita Zavelani-Rossi[1,2], Stefano Longhi[1,2], Guglielmo Lanzani[1,3], Francesco Tassone[3]

1 Dipartimento di Fisica, Politecnico di Milano, P.zza L. da Vinci 32, 20133 Milano, Italy
2 Istituto di Fotonica e Nanotecnologie, Consiglio Nazionale delle Ricerche, P.zza L. da Vinci 32, 20133 Milano, Italy
3 Center for Nano Science and Technology@PoliMi, Istituto Italiano di Tecnologia, Via Giovanni Pascoli, 70/3, 20133 Milano, Italy

*francesco.scotognella@polimi.it

**Abstract:**

Heavily-doped semiconductor nanocrystals characterized by a tunable plasmonic band have been gaining increasing attention recently. Herein, we introduce this type of materials focusing on their structural and photo-physical properties. Beside their continuous-wave plasmonic response, depicted both theoretically and experimentally, we also review recent results on their transient ultrafast response. This was successfully interpreted by adapting models of the ultrafast response of noble metal nanoparticles.

## 1. Introduction

The optical features exhibited by metallic nano-systems have been the subject of an intensive theoretical and experimental research, resulting in applications in several fields, from surface-enhanced spectroscopy, to biological and chemical nano-sensing [1]. Such developments are primarily due to the localized surface plasmon resonance (LSPR), which results in intense optical absorption and scattering as well as sub-wavelength localization of large electrical fields in the vicinity of the nano-object [2-10]. The plasmonic response of these metallic nanostructures is strongly dependent on the type of metal of which they are made, on the dielectric function of the surrounding medium, on the particle shape and, even if to a lesser extent, on the particle size. Nanoparticles of several metals, such as Ag, Au, Cu and Pt, exhibiting plasmonic response in the ultraviolet and in the visible regions of the spectrum have been successfully synthesized in the past decades [11-15]. Elongated metallic nanoparticles have been shown to exhibit plasmonic response in the near infrared, due to excitation of the longitudinal plasmon mode [16-18]. It has been recently reported that direct optical excitation of the metal by intense femtosecond laser pulses induces an ultra-fast modulation of the plasmonic resonance, which can be used for ultra-fast modulation of signals [19], paving the way to *ultrafast active plasmonic*s. A quantitative investigation of such dynamical features is therefore crucial for the development of a new generation of ultra-fast nano-devices. Pump-probe spectroscopy, giving access to the electronic excitation and subsequent relaxation processes in the material, is to date the most suitable tool for the experimental study of the ultrafast dynamical features exhibited by plasmonic nanostructures. So far, several noble metal structures have been investigated, including spherical nanoparticles [20-22] and nanorods [23], revealing that the physical processes of excitation and relaxation are actually similar to those observed in bulk (i.e. in thin films) metallic systems, that are the electron-electron, the electron-phonon and the phonon-phonon interactions [24,25].

In addition to "traditional" metal plasmonic materials, a consistent effort has been made in the last years in the synthesis of heavily-doped semiconductor materials so as to achieve metallic behaviour and plasmonic response in desired wavelength ranges which are not easily accessed by metal nanoparticles. In this respect, copper chalcogenides, such as sulfides and selenides are an interesting class of materials, since a wide copper/chalcogenide stoichiometric ratio is accessible in the synthesis. When this ratio is just below 2:1, these materials express a large number of copper vacancies, which give rise to a large number of holes in the valence band, hence into a "self-doping". This self-doping produces a quasi-metallic behavior of the material. Indeed, Gorbachev and Putilin [26] reported a plasmon band in the reflectivity of p-type copper selenide and copper telluride thin films in the infrared (IR) spectral region. They also found a strong temperature dependence of the spectral position of this band. This behaviour was attributed to a strong dependence of the hole effective mass on temperature.

Recently, first reports on copper chalcogenide nanoparticles also reported about an IR absorption band from oxidized $Cu_2Se$ nanoparticles. However, the particles were embedded in a glass matrix, and such band was ascribed to transitions of electrons trapped in deep states in the matrix [27,28]. Colloidal nanocrystals of copper chalcogenides, mainly $Cu_2S$ and $Cu_2Se$, have been successfully synthesized [29-36]. As in the bulk, also in these nanocrystals Cu can exist in stoichiometric ratios considerably lower than 2:1 with respect to S or Se, i.e. the material may be strongly "self-doped". In 2009, Zhao et al. [29] prepared $Cu_{2-x}S$ nanocrystals with different values of x and clearly demonstrated how the variation from the stoichiometric ratio had a strong influence on the plasmonic response. In 2011, Luther et al. [34] and Dorfs et al. [36] reported detailed studies on the plasmonic properties of $Cu_{2-x}S$ [34] and $Cu_{2-x}Se$ [36] nanoparticles. They investigated the strong influence of self-doping on the plasmonic response of the nanoparticles, and in particular, on their tunability. Recently, the plasmonic response in heavily doped metal and transition metal oxides nanocrystals has also been described [37,38].

While ultrafast plasmonics has been studied extensively in noble metal nanoparticles, not much is known instead about the recently developed heavily-doped semiconductor nanocrystals that likewise exhibit localized surface plasmon resonance. In this Colloquium paper, we first review the pioneering studies of plasmon resonance in heavily doped semiconductor thin films. Then, we report on the chemical synthesis and structural properties of heavily doped semiconductor nanocrystals. Their plasmonic response under continuous-wave optical excitation is illustrated both theoretically and experimentally. Finally, we review the most recent results on the transient (i.e. nonlinear) plasmonic features exhibited by chalcogenide nanocrystals under excitation with ultra-fast optical pulses, including a "gold-like" theoretical model. This model turned out to provide sufficient insights into such first experiments on heavily-doped plasmonic nanoparticles.

## 2. Bulk Chalcogenide Semiconductors

It is worth noting that the metallic behaviour of heavily-doped semiconductors has been reported and extensively studied almost forty years ago in bulk materials and films. Among the different *self-doping* compounds, it is very useful for our argumentation the study of Copper chalcogenides, i.e. sulphide, selenide and telluride. In 1967, Abdullaev et al. described the preparation of $Cu_2Se$ single crystals [39]. They reported the degeneration of the hole gas in these crystals that seemed to be due to structural imperfections of $Cu_{2-x}Se$ introduced during the chemical synthesis, as confirmed by X-ray analysis. This study is one of the first evidences that $Cu_{2-x}Se$ behaves as a p-type degenerate semiconductor with a partially filled valence band.

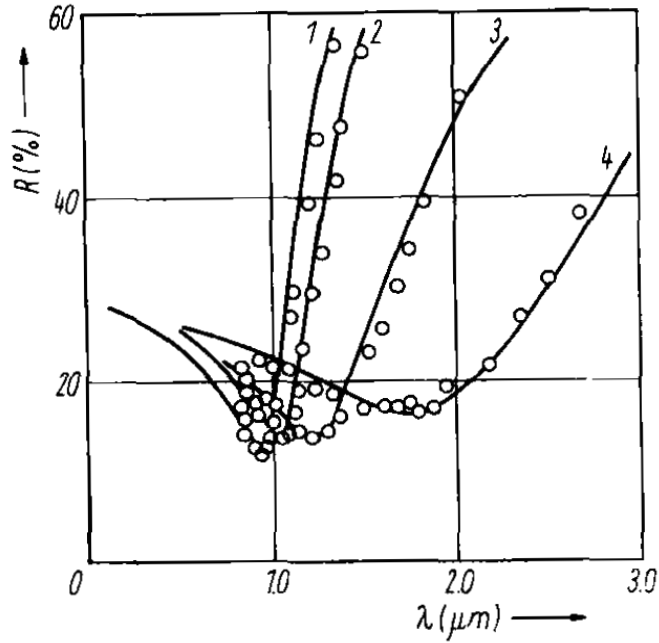

**Figure 1.** Temperature dependence of the reflection spectrum of copper selenide (carrier concentration $N_C$= 2.04 × $10^{21}$ cm$^{-3}$) in the region of the plasma minimum. (1) 100 K; (2) 300 K; (3) 370 K; (4) 450 K. Reprinted with permission from ref. [26].

The plasma resonance of $Cu_{2-x}Se$ and $Cu_{2-x}Te$ have been reported in a pioneering work by Gorbachev and Putilin in 1972 [26], a paper devoted to the determination of the physical parameters of the band structure in these materials. Figure 1 shows the temperature dependence of the reflectance in the region of the plasma minimum for a copper selenide sample with carrier concentration $N_C$= 2.04 × $10^{21}$ cm$^{-3}$. With temperature rising from 100 K to 450 K, the position of the plasma minimum shifts to the longer wavelength region of the spectrum. A similar behaviour in terms of temperature dependence of the plasma minimum in the reflection spectrum was observed for copper telluride [26].

Since the plasma frequency $\omega_P$ is related to the free carriers effective mass $m_s^*$ by the equation

$$\omega_P = \sqrt{\frac{N_C e^2}{m_s^* \varepsilon_0}}, \quad (1)$$

the observed temperature dependence was ascribed to a temperature dependence of the carrier effective mass, according to the following phenomenological formula:

$$m_s^* = m_s^*(T) = (aT^b + c)m_e, \quad (2)$$

In above equations, $T$ is the absolute temperature of the lattice, $e$ is the electron charge, $m_e$ is the free electron mass, $\varepsilon_0$ is the dielectric constant in the vacuum, and $a$, $b$ and $c$ are fitting parameters (see Fig. 2). The origin of such temperature dependence has been attributed to either polymorphic transformations at high temperatures inherent in these materials, or to the complex structure of the valence bands [26,40]. Due to the contribution of the d-states of Copper to the valence-band states, and to the intrinsic positional disorder in the crystal, the calculation of the band-structure in these materials is complex, and it has been possible only recently to obtain a preliminary understanding. In particular, stoichiometric $Cu_2S$ in both cubic and hexagonal phases shows a small but negative band-gap (in the sense that the conduction band crosses the valence band) even at the GW level [41]. However, it is well established that disorder in the positions of the Cu atoms occurs in the real crystals. When this disorder is introduced also in the calculations, which thus require large computational resources not accessible until recently, a small gap is produced. Self-doping further enlarges the gap, both through a Burstein-Moss type of shift, but also through a further enhancement of the band separation related to a narrowing of the valence band (thus presumably also leading to a slight increase of the hole effective mass).

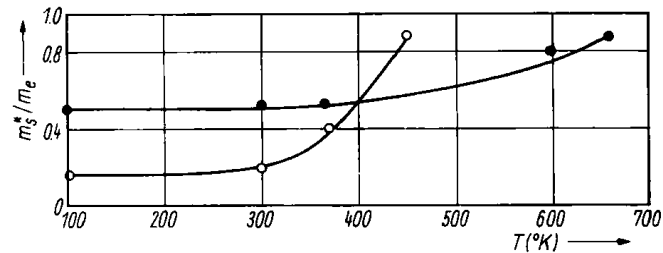

**Figure 2.** Temperature dependence of the effective carrier mass in copper selenide (white circles) and telluride (black circles) as provided by Eq, (2) by fitting *a*, *b* and *c* parameters onto reflectance experimental data. Reprinted with permission from ref. [26].

While calculations at this required level of complexity have not been extended to selenides and tellurides, we expect that most of the very basic conclusions still hold. In particular, the gap is expected to be even smaller, so that its opening presumably is also related to the partial disorder in the location of the copper ions in these compounds.
The relation of the optical gap with the band structure is still controversial, given that only recently more reliable calculations of the band-structure in some of these copper chalcogenides could be carried out. Typically, it is assumed that these materials show an indirect band-gap, mainly based on the square-root energy dependence of the absorption coefficient, as found by Al Mamum *et al.* [42,43] for $Cu_{2-x}Se$ and [44] for $Cu_2S$. However, the magnitude of this coefficient is very large, and comparable to that of direct-gap semiconductors. Indeed, a direct optical gap was found in the calculations for $Cu_2S$, although the complex energy dependence of the optical absorption could not be calculated, certainly due to the large size of the computation related to the disordered structure.

A general overview of the complex phase diagram of $Cu_{2-x}Se$ is reported by Korzhuev in 1998 [45], and is shown in Figure 3 (a). In particular it is clear that around room temperature, and in a wide range of stoichiometries, there are several different stable phases. This complex behaviour is even more evident in $Cu_{2-x}S$ as depicted in a detailed phase diagram around room temperature and small x shown in Figure 3 (b) from Ref. [46]. The different phases are expected to exhibit also different behaviour of the plasma response, thus making difficult a quantitative understanding and application of the physical phenomena.

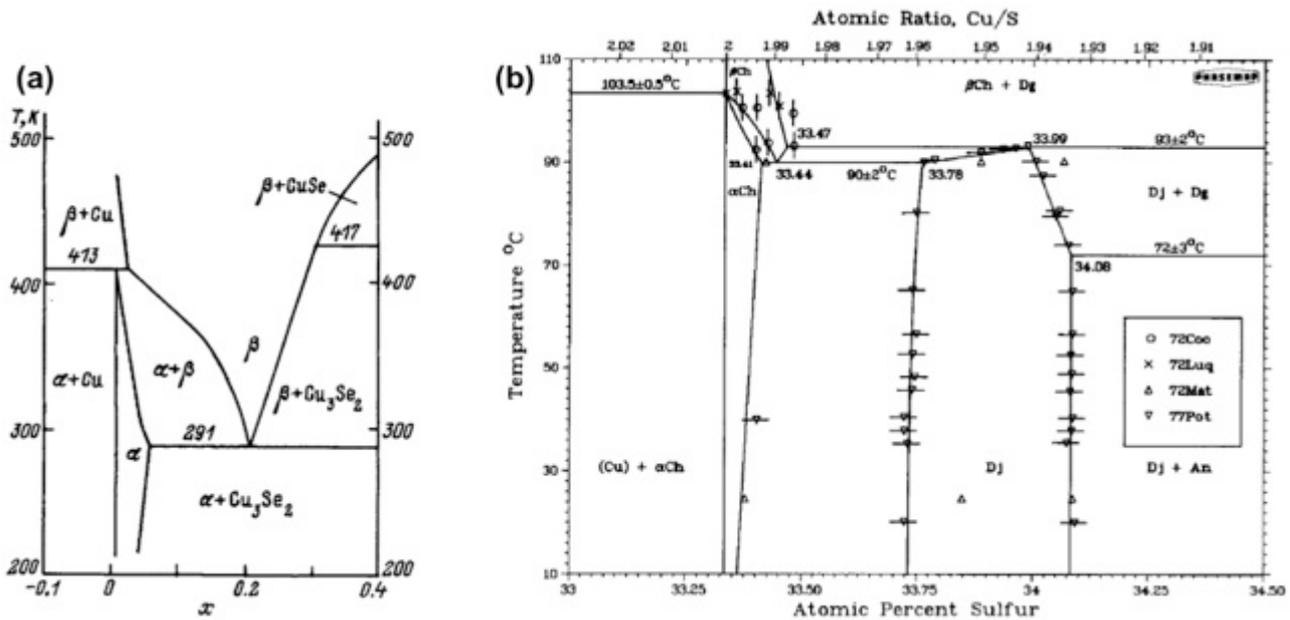

**Figure 3**. (a) Phase diagram of bulk Cu-Se system in the $Cu_{2-x}Se$ compositional region characterized by -0.1<x<0.4, according to [45]. α and β are the α and β chalcocite phases. (b) Detailed phase diagram of $Cu_{2-x}S$ in the region -0.05<x<0.1 as reported in [46]. Here Dj is djurelite, and Dg is digenite. Reprinted with permission from refs [45,46].

Neutron diffraction measurements performed on $Cu_{1.75}Se$ and $Cu_{1.98}$ Se samples for both α- and β-phases by Danilkin *et al.* [47-49] show the dependence on the phase of the kinetics in the copper diffusion in the disordered crystal.

All these pioneering studies evidenced a complex scenario in terms of crystalline phases and subsequently optical and plasmonic response. In particular, the strong dependence of the plasma frequency on temperature is likely to pose severe limitations to the application of these materials as thin films. Instead, wet synthesis of nanoparticles has shown the possibility to realize them in the relatively more controlled crystalline phases, frequently also over a wide range of stoichiometries. This is promising for a robust engineering of plasmonic properties, and successful application of these materials.

## 3. Chemical synthesis and structural properties of heavily-doped nanocrystals

In 2003 Zhao *et al.* [29] reported one of the first wet syntheses of $Cu_{1.8}S$ nanoparticles and additionally investigated on the synthesis of $Cu_{2-x}S$ nanoparticles at different *x* using three different chemical methods: sono-electrochemical, hydrothermal and solvent-less thermolysis. They observed that using the three different approaches, they could obtain nanocrystals with the same value of x and the similar crystalline phase.

It has been demonstrated the possibility to control the size in the synthesis of colloidal copper sulphide nanocrystals ranging between 2 and 6 nm in diameter [18,19,24]. In Figure 4 the transmission electron microscope images of these nanocrystals are shown aside with the electron diffraction patterns (see the insets) and size distribution statistics (see bottom panels).

It is worth noting that these structures can also be assembled into large super-lattices, which could be convenient in devices. Despite a red shift of the plasmon resonance is observed, in parallel with the red-shift of the optical absorption at the band-edge, the plasma response remains well defined and may be expected to have interesting applications [50].

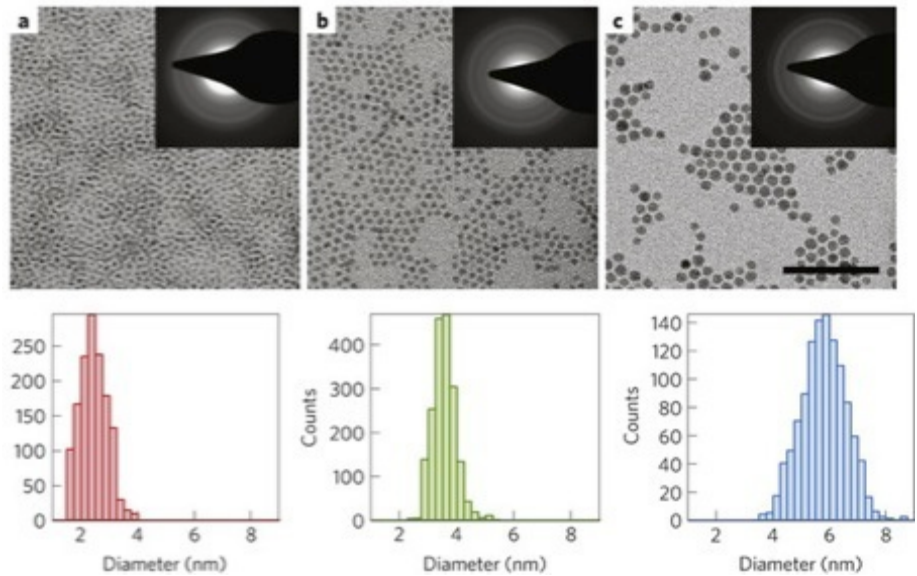

**Figure 4.** Transmission electron micrographs and (insets) electron diffraction patterns (top), and size distribution histograms (bottom) of three QD samples of colloidal copper sulphide with average size of (**a**) 2.4±0.5 nm (**b**) 3.6±0.5 nm, and (**c**) 5.8±0.8 nm. Scale bar represents 50 nm in all images. Reprinted with permission from ref. [25].

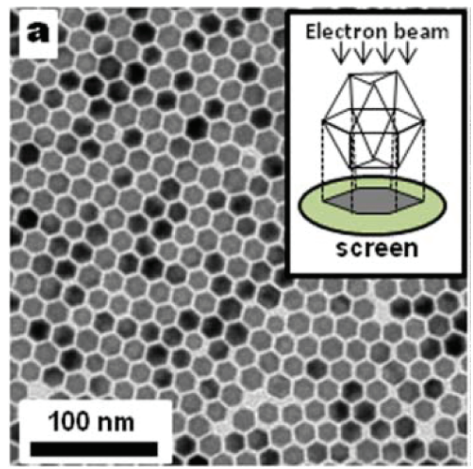

**Figure 5.** (a) TEM image of size-selected $Cu_{2-x}Se$ nanocrystals, having an average size of 16 nm (the size estimated by XRD was 18 nm). The inset shows a sketch of the hexagonal projection of a cuboctahedron shape. Reprinted with permission from ref. [51].

Colloidal copper selenide nanocrystal synthesis has been described in many reports, including work from Deka *et al.* [51], Dorfs *et al.* [36] and Kriegel *et al.* [52]. In Figure 5 TEM images of the synthesised copper selenide nanocrystals are reported.

In literature there are also reports on the synthesis of heavily-doped metal oxide nanocrystals, such as indium tin oxide, aluminium zinc oxide and tungsten oxide. In 2008, Gilstrap *et al.* [53] obtained non-agglomerated indium tin oxide nanocrystals with an average diameter of 5 nm (Figure 6), while Choi *et al.* [54] synthesised monodisperse indium tin oxide nanoparticles with a size ranging from 3 and 9 nm. In 2011, Garcia *et al.* obtained ITO nanoparticles with size ranging from 4 to 12 nm, that were also processed to make electrically conductive thin films [37].

Synthesis and optical properties of AZO NCs have been reported by various groups [55-61]. In particular, Buonsanti et al. proposed a synthesis route to high quality AZO NCs, with sizes that can be varied between 5 and 20 nm, which allow to tune independently the crystal average size and the doping level [60].

Concerning tungsten oxide, there are many known stoichiometrically stable oxygen deficient bulk phases [62], also changing as a function of temperature. In Ref. [62], a transition to metal conductivity is also shown to occur for x > 0.1 in $WO_{3-x}$. This transition to a metallic behaviour also results into a plasmonic optical response in the reflectivity of thin films [63]. Only few of these phases have been investigated at the nanoscale. Moreover, the synthesis of nanoneedles, nanowires and nanorods have been demonstrated [38,64,65].

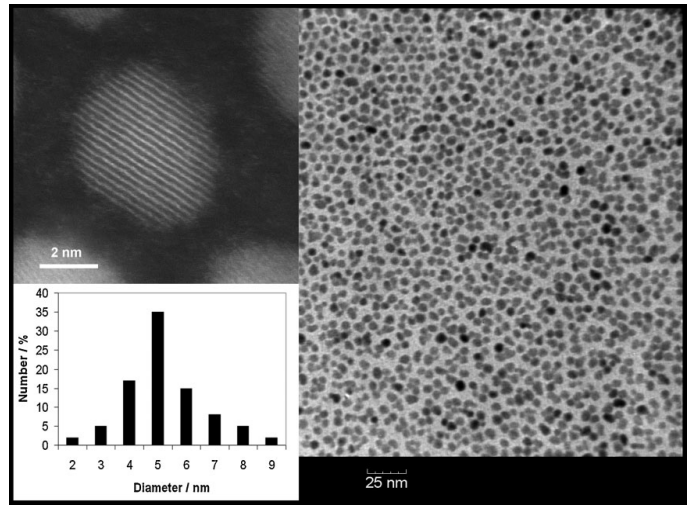

**Figure 6.** TEM picture of a dispersion of indium tin oxide nanoparticles. Upper left inset: STEM high resolution image of a single particle. Lower left inset: size distribution sampling 300 nanoparticles. Reprinted with permission from ref. [53].

## 4. Localized plasmonic resonance in nanocrystals

Localized plasmon resonances in noble metal nanoparticles with radius $R<<\lambda$ can be theoretically described by quasi-static approximation of Mie scattering theory, providing for the absorption cross-section $\sigma_A$ the following expression (see e.g. [66] and references therein):

$$\sigma_A = 4\pi k R^3 \cdot \text{Im}\left\{\frac{\varepsilon(\omega)-\varepsilon_m}{\varepsilon(\omega)+2\varepsilon_m}\right\} \tag{3}$$

where $\varepsilon(\omega)$ is the metal dielectric function at the optical frequency $\omega$ and $\varepsilon_m$ is the dielectric constant of the surrounding medium (assumed to be a linear homogenous isotropic dielectric). The optical extinction of ultra-small particles is dominated by absorption, but for not too small particles a first order correction to quasi-static approximation ought to be included to account for a scattering contribution, with a scattering cross-section given by the following expression [66]:

$$\sigma_S = \frac{8}{3}\pi k^4 R^6 \cdot \left|\frac{\varepsilon(\omega)-\varepsilon_m}{\varepsilon(\omega)+2\varepsilon_m}\right|^2 \tag{4}$$

In above equations, $k=n_m\omega/c$ with $n_m=(\varepsilon_m)^{1/2}$ the refractive index of the dielectric environment and $c$ is the speed of light in vacuum.

The dielectric function of the metal typically entails two different terms, one from the free-carriers in the conduction band (Drude term), $\varepsilon_{\text{Drude}}(\omega)$, and one from the bound electrons which are responsible for interband optical-transitions (Lorentz term), $\varepsilon_{\text{Lorentz}}(\omega)$ [66]. For noble metals, the Drude contribution is dominating in the red and infra-red, where the Lorentz contribution can be assumed to be almost constant, i.e. $\varepsilon_{\text{Lorentz}}(\omega) = \varepsilon_\infty - 1$, with $\varepsilon_\infty$ the high-frequency limit of the material dielectric function. Therefore, it is commonly assumed for gold and silver nanoparticles in the red/infra-red a complex dielectric function given by $\varepsilon(\omega) = \varepsilon_{\text{Drude}}(\omega) + \varepsilon_\infty - 1 = \varepsilon_1(\omega) + i\varepsilon_2(\omega)$, where:

$$\varepsilon_1 = \varepsilon_\infty - \frac{\omega_p^2}{\omega^2 + \Gamma^2}$$
$$\varepsilon_2 = \frac{\omega_p^2 \Gamma}{\omega(\omega^2 + \Gamma^2)} \tag{5}$$

In above equations, $\omega_P$ is the plasma frequency of the free carriers of the system, and $\Gamma$ is the free carrier damping (i.e. the inverse of the carrier relaxation time).

The localized plasmon resonance exhibited by noble-metal nanoparticles is manifested as a pronounced peak in the total optical extinction spectrum $\sigma_E(\omega) = \sigma_A(\omega) + \sigma_S(\omega)$. Since for noble-metals below the plasma frequency $|\varepsilon_2(\omega)| << |\varepsilon_1(\omega)|$ an approximated value for the resonance frequency $\omega_R$ is provided by the so-called Frölich condition: $\varepsilon_1(\omega_R) = -2\varepsilon_m$. Note that in the red/infra-red, where the real part of the metal dielectric constant is monotonically increasing as a function of the optical frequency, the Frölich condition directly implies the red-shift (blue-shift) mechanism of the plasmonic resonance with an increase (decrease) of the refractive index of the surrounding environment.

The quality factor $Q$ (given by the ratio of the surface plasmon resonance bandwidth to the resonance frequency) in the quasi-static limit is known to be independent of particle shape and size and turns out to be ultimately limited by the Ohmic losses experienced by the free-carriers, according to the following equation [67]:

$$Q = \frac{\omega_R}{2\varepsilon_2(\omega_R)} \left. \frac{d\varepsilon_1(\omega)}{d\omega} \right|_{\omega_R} \tag{6}$$

Typically, the quasi-static quality factor $Q$ is limited to few tens for gold nanoparticles and to 50-100 for silver nanoparticles, and retardation effects in larger nanoparticles typically cause the quality factor to decrease below the quasi-static limit because of the presence of scattering (that is radiation) losses. Though it has been demonstrated that the $Q$ factor of an individual plasmonic structure can be sometimes increased beyond the quasi-static limit of Eq. (6) when including the wave retardation for plasmonic nanoparticles with magnetic-dipole response [68], the limited value of the quasi-static $Q$ poses some limitations to plasmonics in sensing applications with ultra-small particles. It is thus envisaged that the development of novel plasmonic materials can provide an enhancement of the quasi-static $Q$ as compared to metallic plasmonic nanostructures.

Localized plasmon resonances with identical features as those observed in noble-metal nanoparticles have been reported in copper calchogenide nanocrystals, and in other non-metallic media, including tin and zinc oxides.

## 4.1. Chalcogenide nanocrystals

Among copper chalcogenide nanocrystals, the $Cu_{2-x}Se$ nanocrystals have been attracted particular interest, since the copper stoichiometry can be accurately controlled upon oxidation/reduction processes as described in detail by both Dorfs et al. [36] and Kriegel et al. [52]. The optical response exhibited by these nanocrystals under continuous wave excitation with near infrared light is plasmonic in nature and can be quantitatively interpreted according to the quasi-static approximation of the Mie theory described above. As example, for $x=0.15$, $Cu_{1.85}Se$ nanocrystals dissolved in toluene are characterized by a relatively narrow plasmonic resonance (about 0.56 eV), with an intense peak around 1050 nm, as shown in Figure 7, which reports the optical extinction spectrum of a solution of $Cu_{1.85}Se$ nanocrystals with estimated particle concentration $N = 13 \times 10^{13}$ cm$^{-3}$, placed in a 0.5 mm thick quartz cuvette. Figure 7 also reports the total extinction cross section $\sigma_E$ of $Cu_{1.85}Se$ nanoparticles computed from Eqs. (3)-(5) with parameters $e_\infty = 10$, $\omega_P = 6.76 \times 10^{15}$ rad/s, and $\Gamma = 6.64 \times 10^{14}$ rad/s [69].

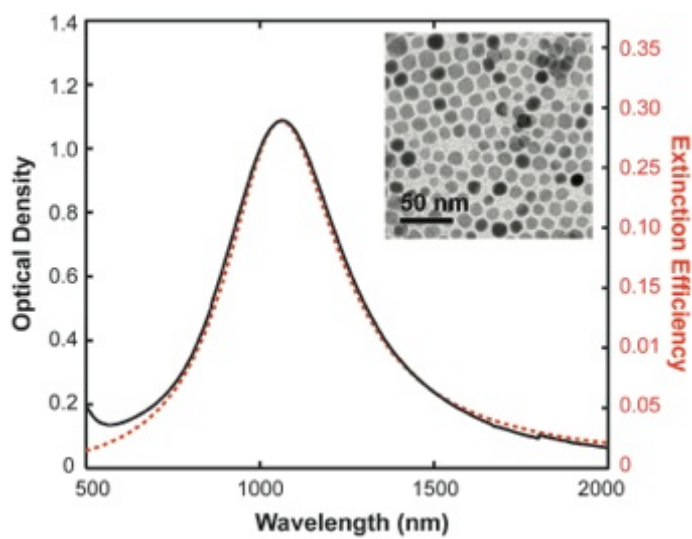

**Figure 7.** Experimental steady-state optical extinction spectrum of a solution of $Cu_{1.85}Se$ nanocrystals dissolved in toluene (solid black line), and theoretically computed extinction efficiency (dashed red line) from Eqs. (3)-(5) (see text for parameters). The inset reports a transmission electron microscopy (TEM) image of the nanocrystals. Reprinted with permission from ref. [69].

The dielectric function $\varepsilon$ depends on $x$, the degree of Cu deficiency, and this results in a plasmon resonance at frequency $\omega_R$ that shows a blue-shift with increasing $x$ [26]. The oxidation process was responsible for the decrease of the Cu:Se stoichiometry from values close to 2:1 down to the lower limit of 1.6:1; an other way to vary $x$ from 0 to 0.4 is via $Ce^{4+}$-based oxidation (Figure 8). In situ elemental analysis in the transmission electron microscope (TEM) assessed the stoichiometry for these two extreme compositions. During this variation in $x$, the optical response in the near-infrared region evolved from a broad band around 1700 nm ($Cu_{1.96}Se$), to narrower band at shorter wavelengths, up to 1100-1150 nm for the most oxidized species. Gradual reduction of the resulting copper deficient nanocrystals by the addition of $Cu^+$ ions restores the optical response of the "as-synthesized" samples. Hence the reduction of *x* to 0, by further addition of Cu(I) salt, led to disappearance of the plasmon band. However, controlled oxidation could restore the absorption in the IR. A control over x in the range from 0 to 0.2 (for oxidation with $O_2$) or from 0 to 0.4 (for oxidation with $Ce^{4+}$) has been achieved, allowing fine-tuning of the NIR absorption band in the 1100-1700 nm range (Figure 9).

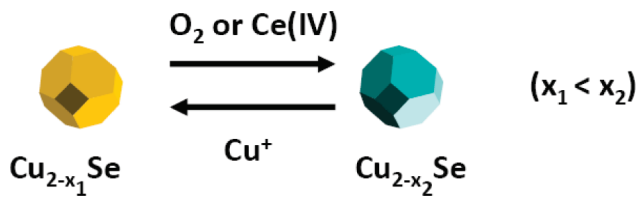

**Figure 8.** Oxidation, which increases copper deficiency, could be carried out either by exposure to air or by stepwise addition of a Ce(VI) complex. Reduction, which decreases copper deficiency, could be carried out by addition of a Cu(I) complex. Reprinted with permission from ref. [36].

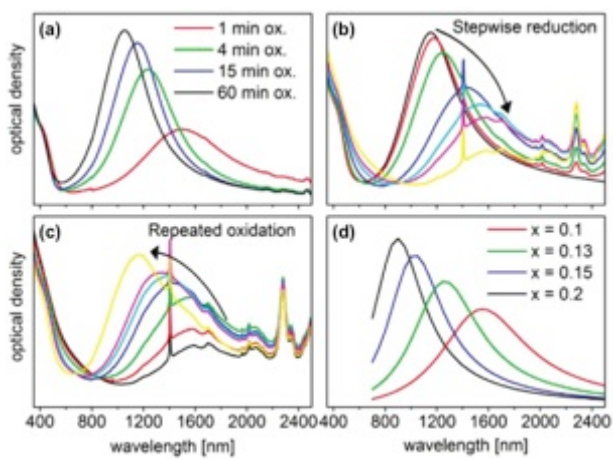

**Figure 9.** (a) Extinction spectrum of the "as-synthesized" $Cu_{1.96}Se$ nanocrystals over time, when they are gradually oxidized, under ambient conditions, to $Cu_{1.81}Se$. (b) Extinction spectra of the oxidized sample ($Cu_{1.81}Se$) when treated with different amounts of a Cu(I) salt. Each step from the black toward the yellow spectrum corresponds to an addition of 5 μL of a 0.02 M $Cu(CH_3CN)_4PF_6$ solution. (c) Evolution of the extinction spectra of the reduced sample from (b) when again oxidized under ambient conditions for 19 h in total from the black to the yellow spectrum. Sharp absorption peaks in panels (b) and (c) are due to methanol and Cu(I) salt absorption. (d) Calculated extinction spectra of single $Cu_{2-x}Se$ nanocrystals for different *x* values. Reprinted with permission from ref. [36].

## 4.2. Localized surface plasmon resonance in other nanocrystals

Several groups have been investigating the optical properties, from the visible to the near infrared, of indium tin oxide (ITO) nanocrystals [70-72]. A plasmonic resonance has been observed, whose position could be tuned from 1600 nm to 2200 nm in two different ways: i) by controlling the concentration of tin dopants as shown in Figure 10) [37,71,73]; ii) electrochemically by applying a bias-voltage [37]. As for the case of other wide band gap n-type semiconductors, the plasmonic band of ITO can be explained via Mie scattering theory using a frequency dependent damping parameter, which depends on scattering from ionized impurities [55,74,75]. However, Wang *et al.* noted that the plasmon resonance appeared only for body-centered cubic (bcc) NCs and not for rhombohedral ones [72].

Another dopant for tin oxide is antimony. In the work by zumFelde et al. it has been investigated the infrared optical properties of antimony-doped tin oxide (ATO) NCs produced by a co-precipitated method [76], where infrared absorption was modulated in intensity by applying a bias voltage.

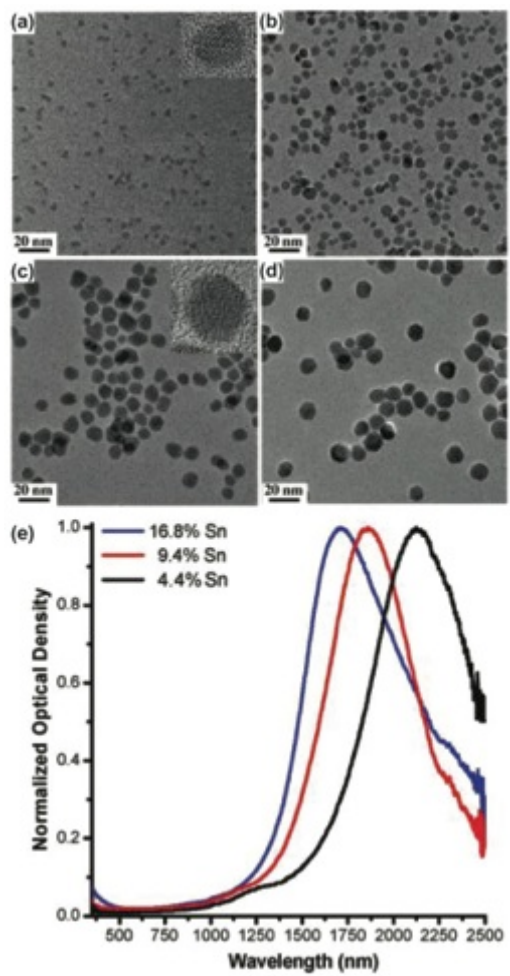

**Figure 10.** Transmission electron microscopy images of ITO NCs of varying size (nm): (a) 4.1 ± 0.6, (b) 7.4 ± 1.4, (c) 10.2 ± 1.7, and (d) 12.1 ± 1.5. (e) Extinction spectra of NCs with varying tin content (%) in solvent dispersions. Reprinted with permission from ref. [26]. [37].

Zinc oxide nanocrystals (n-type semiconductors) show enhanced IR absorption, obtained by charge-transfer doping or by irradiating the crystals with above-bandgap UV light, as observed by Shim et al. [77]. Aluminium doped zinc oxide NCs (AZO) have a clear infrared plasmon peak, paving the way to the use of such material as a cheaper and eco-friendly alternative to ITO NCs for applications in flexible and wearable electronic devices. It has been shown that, by varying the concentration of Al substitutional ions between 0 and 8 %, it is possible to tune the plasmon resonance between 3 and 10 μm (3200 to 1000 cm$^{-1}$), while keeping the NCs transparent in the visible range [60]. Hammarberg et al. [57] reported a synthesis strategy to realize indium doped zinc oxide NCs, while Cohn et al. described a synthesis route in which zinc could be substituted by manganese [78].

An additional way, besides doping, to make ZnO NCs conductive is to photochemically charge them [79]. These NCs have the same optical properties of Al doped ZnO NCs, but show a different chemical reactivity [80].

Manthiram and Alivisatos [38] showed tungsten oxide nanocrystals with potential interest for photo-chemical applications, also owing to their robustness. The electronic band gap of 2.6 eV, suitable for photovoltaics, is tunable by modifying the stoichiometry. Moreover, stoichiometry deficient $WO_{3-x}$ nanocrystals have an intense absorption band between the visible and near infrared region of the spectrum, due to a tunable localized surface plasmon resonance (Figure 11). For instance, they showed that the absorption spectrum of $WO_{2.83}$ NCs could be tuned from 1.38 eV (900 nm) to 1.13 eV (1100 nm), by heating in presence of oxygen and incorporating new oxygen atoms in the crystal lattice, and therefore changing the carrier density.

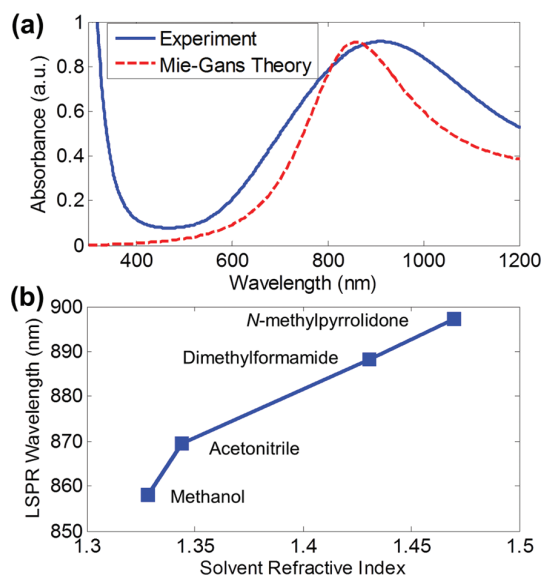

**Figure 12. (a)** Comparison of experimentally observed and theoretically predicted absorption spectra of the short-axis mode of $WO_{2.83}$ rods in N-methylpyrrolidone. **(b)** Experimentally observed short-axis LSPR wavelength as a function of the refractive index of the solvent. Reprinted with permission from ref. [38].

## 5. Ultrafast plasmon response in semiconductor nanocrystals

Very recently, non-metallic plasmonic nanocrystals have attracted an increasing interest as non-linear plasmonic materials. As for the case of static optical response, the dynamical optical features exhibited by nonstoichiometric calchogenide nanocrystals turned out to be very similar to the nonlinear plasmonic response exhibited by noble-metal nanoparticles. Experimental studies conducted by pump-probe technique have been reported by Scotognella *et al.* [69] and by Kriegel *et al.* [52] for compounds including $Cu_{2-x}S$, $Cu_{2-x}Se$ and $Cu_{2-x}Te$ ($x>0$). The pump-probe technique allows for an accurate investigation of the transient optical response of a system after exposure to an intense laser pulse (pump pulse). A weak pulse (probe pulse) launched onto the sample at a controlled time delay with respect to the arrival of the pump pulse is exploited to monitor the temporal evolution of the optical extinction of the sample (usually measured in terms of a differential absorption or a differential transmission). The differential absorption measurements in correspondence to the peak of the localized plasmon resonance of $Cu_{2-x}S/Se/Te$ (Fig. 12) exhibit a common general trend with two well separated temporal dynamics: a short-time dynamics, on the time-scale of few ps, and a long-time dynamics, on the time scale of hundred ps. These distinct temporal behaviours have been attributed to the dynamical features of the two fundamental processes of *carrier-phonon* interaction and *phonon-phonon* interaction respectively [52,69], similarly to what observed in gold and silver nanoparticles. Actually, when the pump pulse is in the infrared region of the spectrum, the initial Fermi distribution of free carriers of the systems is strongly perturbed by pump absorption; the pump pulse creates energetic carriers that are not in thermal equilibrium and within a few hundred femtoseconds a new Fermi distribution is reached via strong carrier-carrier scattering, resulting in a thermalized free carrier gas with higher temperature than the lattice (hot carriers).

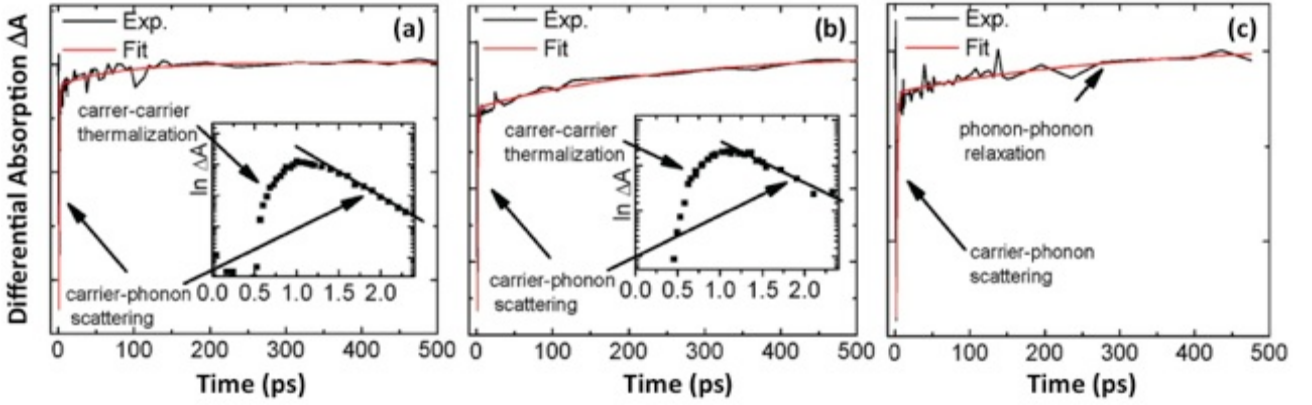

**Figure 12.** Experimental and numerical pump-probe differential absorption measurements in chalcogenide nanocrystals: (a) $Cu_{2-x}S$, (b) $Cu_{2-x}Se$, (c) $Cu_{2-x}Te$. Insets: carrier-carrier thermalization. Reprinted with permission from ref. [52].

Subsequently, within the following few picoseconds the carrier gas cools down by releasing its excess energy to the lattice through carrier-phonon coupling. Ultimately, within hundreds of picoseconds, the nanoparticle releases its energy to the environment, with heat conduction to the surface of the nanoparticles provided by phonon-phonon coupling [35,36].

A first attempt to provide a consistent modeling of the temporal dynamics above described has been reported for $Cu_{2-x}Se$ [69], by mimicking a theoretical model employed for noble-metal nanoparticles. Since the ps dynamics is related to the cooling of the hot carrier gas generated by absorption of the pump energy, the time evolution of the system within ps time scale can be modeled as a result of a heat transfer between a thermalized gas of carriers at temperature $T_C$ and the lattice at a lower temperature $T_L$, via carrier-phonon scattering process. The so-called Two Temperature Model (TTM) quantitatively accounts for such transfer according to the following coupled equations [35]:

$$\begin{cases} \gamma T_C \dfrac{dT_C}{dt} = -G(T_C - T_L) + P_A(t) \\ C_L \dfrac{dT_L}{dt} = G(T_C - T_L) - G_L(T_L - T_0) \end{cases} \quad (7)$$

where $\gamma T_C$ is the heat capacity of the carrier gas, being $\gamma$ the so-called *carrier heat capacity constant*, $C_L$ is the heat capacity of the lattice, $G$ and $G_L$ are the carrier-lattice coupling factor and the lattice-environment coupling factor respectively (to be determined as fitting parameters of the model), $T_0$ is the environmental temperature (which is assumed constant) and $P_A(t)$ is the pump power density absorbed in the volume of the metallic system, that can be estimated from continuous-wave optical measurements (see [69] for details). From numerical solution of Eq. (7) the temperature dynamics induced by the pump pulse is retrieved.

The effects of the carrier and lattice temperatures on the dielectric function depend on the detailed band structure of the material, and on the energy of the probe photon. It is well accepted that Cu$_{2-x}$Se behaves as a p-type degenerate semiconductor with a partially filled valence band (see section 2). According to Al-Mamun *et. al.* [42] (cf. section 2) it is expected that the pump photon at 1.19 eV energy is absorbed by an intra-band process alone similarly to what happens in noble metals in the near infrared, and no indirect interband transitions are expected to occur. The subsequent heating of the carrier gas results in a smearing of the Fermi distribution, and gives rise to a modulation of the inter-band transition probability for the probe light at 1.38 eV and at 0.95 eV [Fig.14(a)], which results in a modulation $\Delta\varepsilon_\infty$ of the $\varepsilon_\infty$ parameter in the dielectric function of Eq. (5).

In metals, it is well known that this modulation is real and proportional to the carrier excess energy [33,36] and thus it scales quadratically with the carrier temperature $T_C$,

$$\Delta\varepsilon_\infty = \eta T_C^2 \tag{8}$$

with $\eta$ being a fitting parameter to be determined.

The heating of the lattice results in a modulation of the $\Gamma$ parameter related to the free carriers, as a consequence of the linear increase of the free carrier scattering with increasing lattice temperature, like in metallic systems [36]:

$$\Gamma = \Gamma_0 + \beta(T_L - T_0) \tag{9}$$

where $\Gamma_0$ is the room temperature carrier damping and $\beta$ a constant parameter, to be estimated from experimental measurements (see [69] for details).

The temperature-dependent dielectric function of Cu$_{2-x}$Se is thus given by:

$$\varepsilon(T_C, T_L) = \varepsilon_\infty + \Delta\varepsilon_\infty(T_C) - \frac{\omega_p^2}{\omega^2 + \Gamma^2(T_L)} + i\frac{\omega_p^2 \Gamma(T_L)}{\omega(\omega^2 + \Gamma^2(T_L))} \tag{10}$$

with $T_C$ and $T_L$ provided by numerical solution of the TTM [see Fig. 13(b)].

The theoretical differential transmission can be then computed as $\Delta T/T = \exp(-\Delta\sigma_E NL) - 1$, with $\Delta\sigma_E$ the variation attained by the extinction cross-section according to the quasi-static formulas of Eq.(3)-(4), and above Eq.(10) for the metal dielectric function.

The theoretical prediction of this semi-empirical model turns out to be in excellent agreement with the experiments on Cu$_{2-x}$Se nanoparticles, as detailed in Figure 14, showing the temporal dynamics of the $\Delta T/T$ probed at 900 nm [Fig. 14(a)] and at 1300 nm [Fig. 14(b)], obtained by exciting the nanocrystals with different pump fluences. At 900 nm, the experimental results revealed $\Delta T/T > 0$ right after the absorption of the pump beam, with a maximum value of about 40% under the maximum pump fluence of 4.45 mJ/cm$^2$, whereas at 1300 nm $\Delta T/T < 0$ was observed, with maximum (negative) value of about -3% under a pump fluence of 1.87 mJ/cm$^2$. In both cases, a monotonic and fast decrease of the signal was then observed within a few ps [Figs. 14(a) and (b)] followed by a much slower (ns time scale) decay [Fig. 14(c) for the signal probed at 900 nm], leading to complete recovery of the initial condition (before pump arrival) within a few ns.

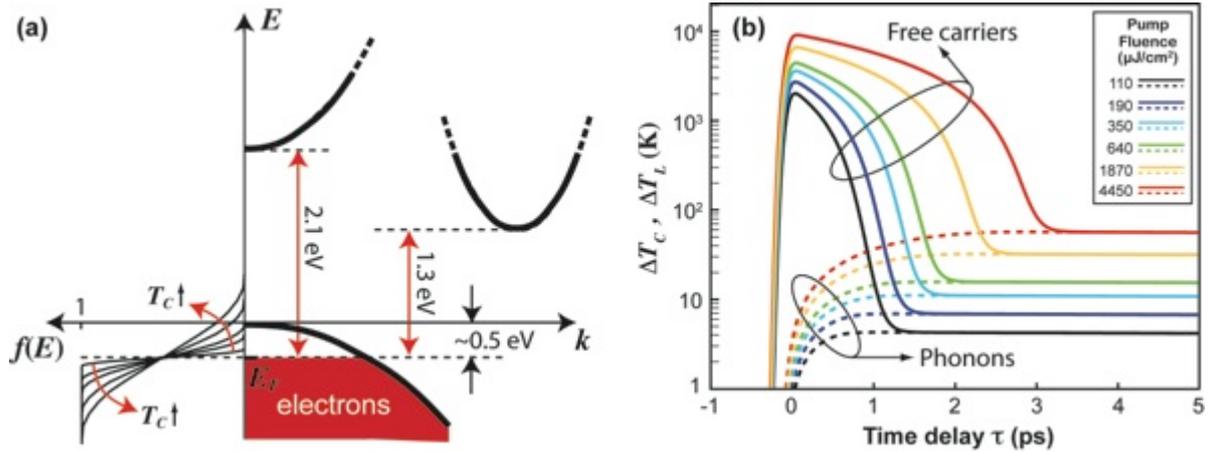

**Figure 13.** (a) Tentative sketch of band diagram in $Cu_{1.85}Se$ for optical transitions in the visible. The smearing of the electron distribution $f(E)$ due to intraband pump absorption and the subsequent carrier temperature ($T_C$) increase is also illustrated. (b) Carrier (hole) and lattice temperature numerically computed from the TTM of Eq. (7). Reprinted with permission from ref. [69].

It is worth noting that the maximum $\Delta T/T$ exhibited by $Cu_{1.85}Se$ nanocrystals exceeds by at least one order of magnitude the $\Delta T/T$ reported in metallic systems for comparable pump fluences (cf. [33,35]). This is attributed to the low carrier density and structural peculiarities of $Cu_{1.85}Se$. In particular, the lower carrier density is responsible for a smaller carrier heat capacity and for a much higher effective carrier temperature at comparable fluences.

Also, at long delays, the significant residual optical response results from a larger dependence of the Drude broadening $\Gamma$ on the lattice temperature compared to metals. Gorbachev and Putilin, for thin films of similar composition and in the same temperature range, found a strong red-shift of the plasma frequency with increasing lattice temperature [27] that was not observed in $Cu_{2-x}Se$ nanoparticles. This shift was attributed to the polymorphic phase transitions of a bulk films $Cu_{2-x}Se$ leading to an increase of the average carrier mass with increasing lattice temperature (cf. section 2). The absence of these phase transitions in $Cu_{2-x}Se$ nanoparticles was confirmed by X-ray powder diffraction measurements in a temperature range between room temperature and 470 K, that clearly indicated that the nanoparticle have cubic phase and no phase transition occurs [69]. Nevertheless, the increase with temperature of the disorder in one of the copper/vacancy sub-lattice is still occurring, and it presumably relates to the large temperature coefficient $\beta$ of the plasma broadening. Indeed, this broadening is certainly dominated by the scattering of carriers on vacancies and other defects.

From the numerical solutions a carrier-phonon coupling factor $G = 1.2 \times 10^{16}$ W m$^{-3}$ K$^{-1}$ has been determined. It must be noted that this estimation is representative of the high temperature range of thousand kelvin explored in the experiments, and provides a $G$ factor one order of magnitude lower than in Au in the same temperature range [66].

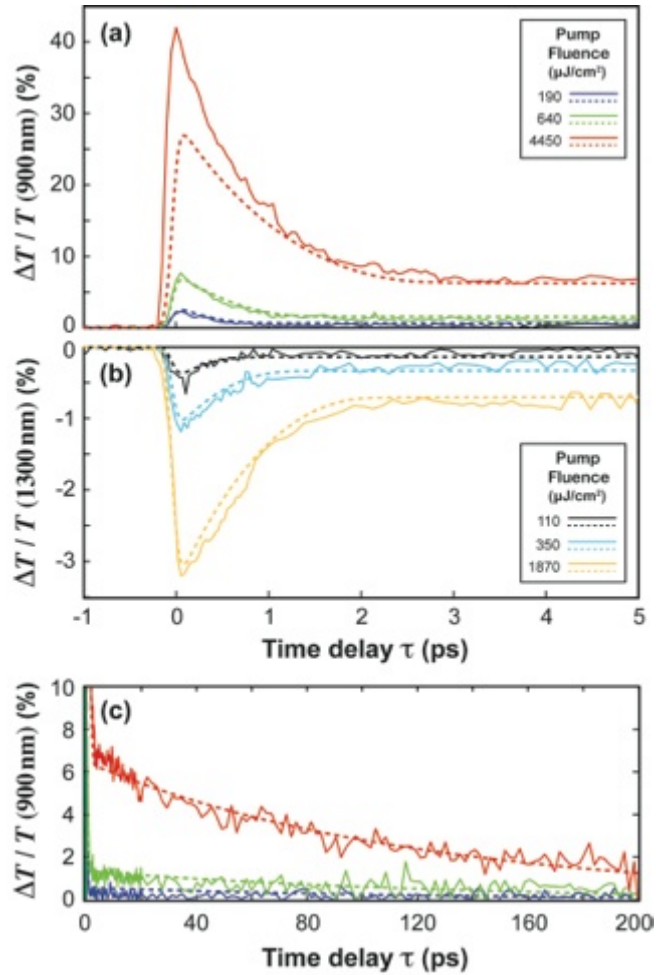

**Figure 14.** Differential transmission signal from a solution of $Cu_{1.85}Se$ nanocrystals dissolved in toluene for short-time dynamics at (a) 900 nm and at (b) 1300 nm probe wavelength, and (c) long-time dynamics at 900 nm. The experimental results (solid lines) are compared with numerical calculations (dotted lines) for the different incident pump fluences. Reprinted with permission from ref. [69].

Since the *G* factor is proportional to the carrier density [81], this lower *G* is in agreement with the lower carrier density in $Cu_{1.85}Se$ compared to Au. It has been also estimated $\eta = 7 \times 10^{-9}$ $K^{-2}$, [see Eq. (8)], which is about five times lower than what is found for silver structures (films) at 900 nm probe wavelength [36].

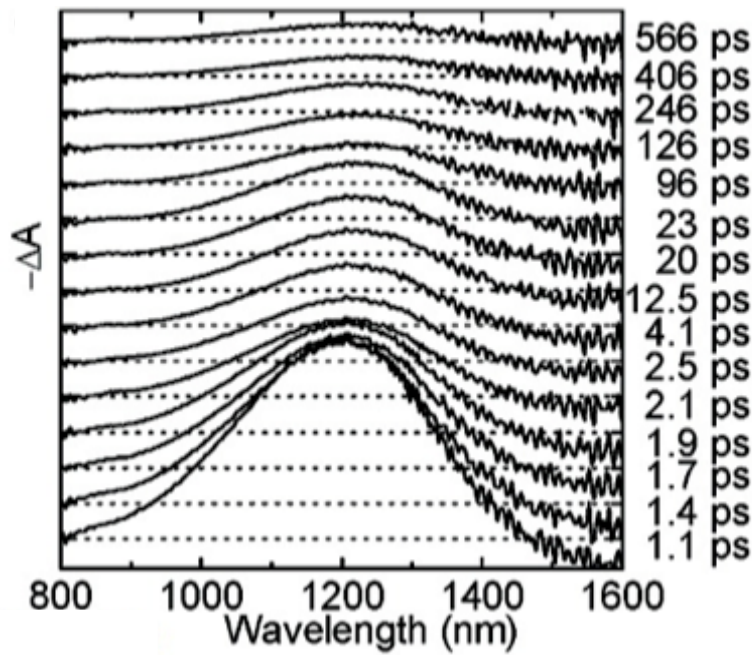

**Figure 15**. Transient absorption spectra at different time delays of $Cu_{2-x}Se$ nanocrystals (with $x>0$) pumped with 100 fs pulsed amplified laser at 400 nm. Reprinted with permission from ref. [52].

A further development of the presented investigation should entail a spectral analysis of the transient plasmonic response, both in terms of experiments and modeling. A preliminary experimental result is available from Kriegel et al. [52], as reported in Fig. 15, showing a derivative feature of the plasmonic transient absorption, in line with the gold-like model introduced in Ref. [69]. At the present point, it is not however possible to draw a quantitative comparison. In particular it is not evident whether the peak broadening is changing in time, as would be expected from its dependence on the lattice temperature [69]. A similar spectral plasmon response was also reported in ref. [52] for heavily doped copper sulphide, even if in the experimental results the derivative feature is less pronounced due to the much larger width of the plasmon peak, comparable to the instrumental spectral window.

**Conclusions**

We reviewed the very recent developments attained in the chemical synthesis, optical characterization and theoretical modeling of heavily-doped semiconductor nanocrystals as novel plasmonic nanoparticles. The capability to control the free-carrier density of the material is paving the way to a new approach in plasmonics, with the potential to overcome some fundamental limitations related to the physics of noble-metals. In particular it has been demonstrated that non-stoichiometric chalcogenides and tin, zinc and tungsten oxides offer the possibility to tune the plasmonic resonance of ultra-small nanoparticles in a broad wavelength range. Also, some of these novel synthetic plasmonic media (chalcogenides) exhibit an optical nonlinearity with almost identical features as that reported in gold and silver, that is a giant incoherent third-order nonlinearity of thermal nature. Most interestingly, thanks to the much lower carrier density as compared to noble metals, this nonlinearity turned out to be even stronger than in gold and silver. It is thus envisaged that the new route of nonlinear plasmonics, recently opened by some pioneering works (see [82] and references therein) can strongly benefit from these novel materials, for the development of ultra-fast all-optical switching devices as well as nonlinear nanosensors.

Though being still at a pioneering level, the proposal and development of non-metallic plasmonic media have already demonstrated the potential of a breakthrough in plasmonics and nonlinear optics. It is worth noting that despite many of these compounds are dated to the seventies as bulk materials, like chalcogenides, it is the very recent capability of nanochemistry that opened a new way for their application into optics and plasmonics, in view of the more versatile synthetic route as compared to more traditional techniques to produce bulk thin films, involving both vacuum and high temperatures. Furthermore, the wet chemistry approach gives direct access to bottom-up nanostructuring with subsequent tailoring of the localized plasmon features. In particular, nanoparticle shaping (in rods, plates, etc.) is another ingredient that can further increase the interest on these new materials by providing the geometrical degrees of freedom explored in the noble-metal nanostructures to tailor the features of plasmonic resonances and eventually profit from retardation effects, leading as example to plasmonic-nanoantennas with enhanced nonlinearity thanks to an engineered plasmonic medium.

We believe that the progress in the field will be mostly subjected to the capability of providing a more consistent and detailed comprehension of the optical and plasmonic properties exhibited by these novel structures under light excitation with ultrafast laser beams. In particular, the role of non-thermalized carriers is still unexplored, and new insight is expected from a more accurate investigation of the plasmon dynamics by means of broad-band pump-probe experiments with few fs resolution time. Also, the effect of quantum confinement in plasmonic nanocrystals of few nm radius can be enhanced by the low carrier density, and carriers surface scattering is expected to be more prominent too as compared to gold or silver nanoparticles of similar size. Size effects on the plasmonic response in heavily-doped nanocrystals have been so far disregarded.

The family of heavily-doped semiconductor nanocrystals is also expected to grow in the future and eventually provide novel plasmonic media operating in the visible with reduced ohmic losses, and thus better quality factors in the quasi-static limit of ultra-small nanoparticles. The latter circumstance is mostly envisaged to improve the figure of merit of nano-sensors based on ultra-small plasmonic nanoparticles. Also, the capability to tailor the dielectric constant of the plasmonic medium represents a new degree of freedom for the engineering of plasmonic metamaterials [83,84] that can be exploited at the level of an individual meta-atom.

*Acknowledgements*
F. Scotognella acknowledges financial support from the project FP7-ICT-248052(PHOTOFET). A. R. Srimath Kandada acknowledges the project PITNGA-2009-237900(ICARUS). G. Della Valle and S. Longhi acknowledge financial support from The Fondazione Cariplo through the research project entitled "New Frontiers in Plasmonic Nanosensing" (contract N. 2011-0338). The authors thank Alberto Comin and Liberato Manna for helpful discussions.